# Photonic skin-depth engineering

**Saman Jahani and Zubin Jacob[*]**

*Department of Electrical and Computer Engineering, University of Alberta, Edmonton, Alberta, T6G 2V4, Canada*
*\*Corresponding author: zjacob@ualberta.ca*

Recently we proposed a paradigm shift in light confinement strategy showing how relaxed total internal reflection and photonic skin-depth engineering can lead to sub-diffraction waveguides without metal (S. Jahani and Z. Jacob, "Transparent sub-diffraction optics: nanoscale light confinement without metal," Optica 1, 96-100 (2014)). Here, we show that such extreme-skin-depth (e-skid) waveguides can counter-intuitively confine light better than the best-case all-dielectric design of high index silicon waveguides surrounded by vacuum. We also analytically establish that figures of merit related to light confinement in dielectric waveguides are fundamentally tied to the skin depth of waves in the cladding, a quantity surprisingly overlooked in dielectric photonics. We contrast the propagation characteristics of the fundamental mode of e-skid waveguides and conventional waveguides to show that the decay constant in the cladding is dramatically larger in e-skid waveguides, which is the origin of sub-diffraction confinement. We also propose an approach to verify the reduced photonic skin depth in experiment using the decrease in the Goos-Hanschen phase shift. Finally, we provide a generalization of our work using concepts of transformation optics where the photonic-skin depth engineering can be interpreted as a transformation on the momentum of evanescent waves.

## 1.Introduction

Conventional optical waveguides confine light by total internal reflection inside a core surrounded by a cladding with lower refractive index than the index of the core ($n_{core} > n_{cladding}$) [1]. For dense photonic integration applications, it is desirable to miniaturize the size of such optical waveguides. However, when the core size decreases, light is weakly confined inside the core and decays slowly outside, i.e. the skin depth in the cladding increases. One approach for reducing the skin depth is to enhance the contrast between the refractive index of the core and cladding. However, for isotropic claddings the lowest refractive index material that can be used is air whereas the highest index is that of silicon. Thus there is a fundamental limitation to reduce the size of conventional optical waveguides. This size limitation can be surpassed using metallic claddings [2–4] or ENZ metamaterials [5,6], but due to their high optical losses it is difficult to use them in dense photonic integrated circuits.

There are two widely used all-dielectric strategies for light confinement: photonic crystal and slot waveguides. Photonic crystal waveguides work based on Bragg reflection [7]. The waveguide modes in such designs are not scattered at sharp bends and it can be confined within low index cores. Slot waveguides confine light in a tiny low-index gap surrounded by high index dielectrics [8]. However, none of these all-dielectric confinement strategies are suitable for photonic integration due to the cross talk [9].

We recently showed that if a dielectric waveguide is surrounded by a transparent anisotropic cladding, the first propagating TM mode can be tightly confined inside the core irrespective of core size [10]. The most striking aspect is that the required anisotropy can be achieved by lossless dielectric media (all-dielectric metamaterials). Therefore the propagation length is very long which is one of the most important figures of merit (FOM) for designing nano-waveguides. Simultaneously, we showed that sub-diffraction photonic mode sizes can be achieved and cross-talk between nano-waveguides can be substantially reduced.

In this paper, we introduce the concept of momentum transformation based on the rules of transformation optics to shed additional light on the phenomenon of relaxed total internal reflection to confine light (section 1). We show that momentum transformations can be applied to a waveguide with arbitrarily shaped cross section. Due to this transformation, the guided mode is strongly confined and it is quasi transverse electromagnetic (quasi-TEM). Unlike other waveguides which support TEM modes [11,12], these waveguides do not need two reflectors or perfect conductors (section 2 and 3). We illustrate that even non-magnetic anisotropic claddings can outperform vacuum in term of confinement (section 4). We also explore in full detail the properties of 1D e-skid waveguides (section 5). Our key result is that all figures of merit related to light confinement in waveguides is connected to the skin depth of light in the cladding, a quantity surprisingly un-engineered in dielectric photonics. We calculate the propagation constant and the decay constant dispersion of an e-skid waveguide to show that the skin depth in the cladding is dramatically reduced due to the anisotropy. We present analytical expressions for three well-known figures of merit to compare e-skid waveguides with conventional dielectric slab waveguides in terms of light confinement (section 6). Finally, we show that reducing the skin depth also causes Goos-Hänchen phase shift reduction at the interface, which is useful for verifying the skin depth experimentally (section 7).

## 2. Relaxed total internal reflection

We look for the solution for light confinement using the rules of transformation optics (TO) which state that Maxwell's equations written in a transformed coordinate system preserve their original form if the material parameters are renormalized [13,14]. We introduce the concept of transforming optical momentum- the physical quantity which governs whether a wave propagates or decays in a medium. We emphasize that this approach, in contrast with previous approaches which primarily dealt with propagating waves [14–16], allows control over evanescent waves which is necessary for waveguiding. If a Cartesian mesh in a region of empty space is transformed according to $x' = f(x)$, $y' = f(y)$, $z' = f(z)$ the optical momentum of propagating or evanescent waves in the region is then transformed to

$$\frac{k_{x'}^2}{h_x^2} + \frac{k_{y'}^2}{h_y^2} + \frac{k_{z'}^2}{h_z^2} = k_0^2 \qquad (1)$$

where the coordinate transformation is characterized by the Jacobian matrix diag $[h_x^2, h_y^2, h_z^2]$, the transformed wavevector $\vec{k} = [k_{x'}, k_{y'}, k_{z'}]$ and $k_0 = 2\pi/\lambda = \omega/c$ is the free space wavevector. The optical momentum transformation, on comparison with the dispersion relation for vacuum, is found to be $k_{x'} = h_x k_x$, $k_{y'} = h_y k_y$ and $k_{z'} = h_z k_z$.

We now revisit the conventional light guiding mechanism of total internal reflection at the interface of two dielectrics using momentum transformations. A plane wave travelling in vacuum (region I) is partially reflected back at $x = 0$ because there is a discontinuity in the "electromagnetic grid" representing optical space (Fig. 1). Electromagnetic boundary conditions require the tangential momentum and hence the phase to be continuous across this interface ($k_{z1} = k_{z2}$). For a given wave incident in a particular direction with $\vec{k} = [k_{x1}, k_{y1}, k_{z1}]$, the ray can be completely reflected back if the transformed momentum in the tangential direction $k_{z1}/h_z$ exceeds the maximum possible momentum in the medium ($k_{z1}^2/h_z^2 > k_0^2$) (Fig. 1.a). This causes the wave to decay away along the $x$-direction in region $x > 0$. Since $k_{z1} < k_0$, we arrive at the condition for the possibility of total internal reflection that the transformation should be such that $h_z < 1$. We note that this condition is different from the well-known condition of $n_1 > n_2$ as a condition for total internal reflection of light moving from medium 1 to 2. The condition is in fact reduced to

$$n_1 > n_{2x} \qquad (2)$$

Eqn. 2 holds for the incident angle ($\theta$) greater than $\theta_c = \sin^{-1}(n_{2x}/n_1)$, where $[n_{2x} \quad n_{2z} \quad n_{2z}]$ is the refractive index tensor of the second medium and $x$ axis is normal to the interface. The interface lies in the $y-z$ plane. We termed this condition as relaxed total internal reflection (relaxed-TIR) [10] since it leaves a degree of freedom unexplored: the perpendicular component of the dielectric tensor.

For this set of transformations that cause total internal reflection, the wave extends evanescently into the second medium. Note that the total internal reflection is governed by the momentum transformation only in the $z$ direction and not the $x$ direction. Using this additional degree of freedom, we transform the optical momentum of evanescent waves to lead to enhanced confinement of the wave in the region with $x > 0$ (Fig. 1.b). We choose a transformation that compresses the optical grid along the $x$ direction with $h_x \gg 1$. This increases the momentum of the wave along the $x$ direction and hence causes a faster decay of evanescent waves in region II. Note that this momentum transformation is valid for both polarizations but requires optical magnetism which is difficult to achieve. For non-magnetic media and TM polarized light, we can arrive at an all-dielectric condition

$$n_{2z} \gg 1 \qquad (3)$$

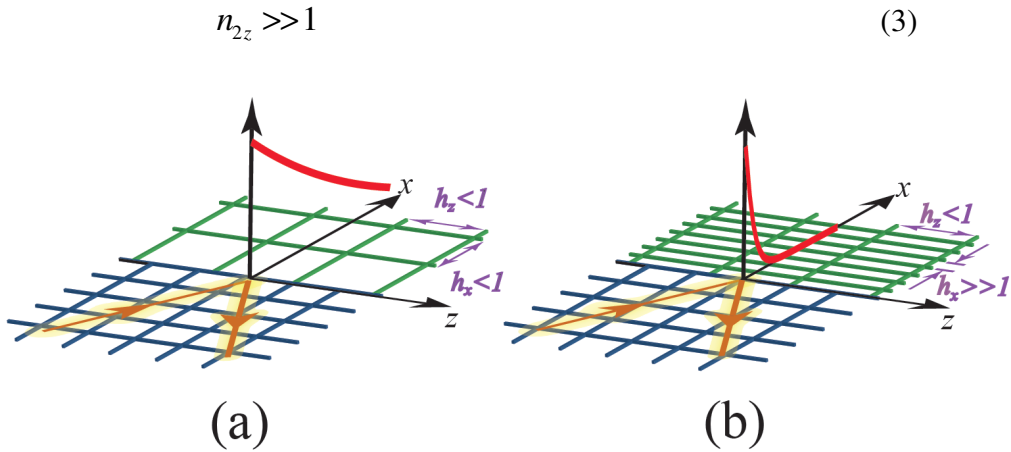

Fig. 1. The phenomenon of refraction and reflection revisited using transformation of optical momentum. Rays of light are reflected and refracted at an interface since the mesh representing electromagnetic space has a discontinuity. (a) Total internal reflection can be viewed as a transformation of optical momentum. When grid sizes in the second medium become large enough, the incident ray is totally reflected and evanescently decays in the second medium. (b) Only one component of the dielectric tensor controls the total internal reflection condition. By transforming the space in the other direction, we can control the momentum of evanescent waves and consequently decrease penetration depth in the second medium.

to increase the momentum of evanescent waves i.e. make them decay faster confining them very close to the interface (Fig 1.b). The skin depth for transparent media at TIR in the second medium is defined as:

$$\delta(\theta) = \frac{1}{k_{2x}} = \frac{n_{2x}}{n_{2z}} \frac{1}{2k_0 \sqrt{(n_1 \sin(\theta))^2 - n_{2x}^2}} \tag{4}$$

which immediately reveals that by increasing $n_{2z}$, we can confine the evanescent wave decay in the second medium ($k_{2x}$) and hence reduce its skin depth ($\delta$).

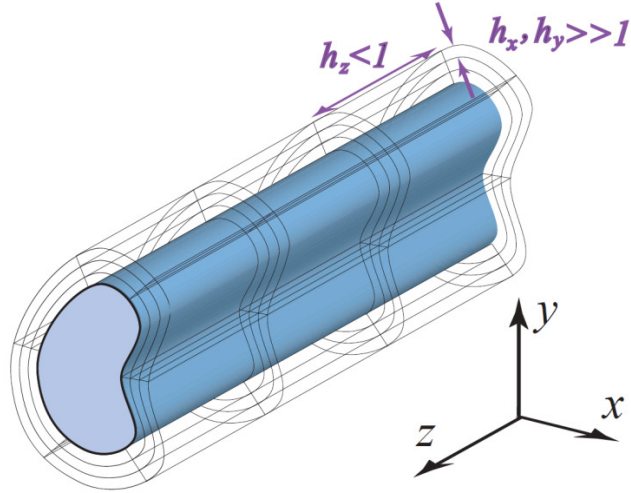

Fig. 2. Light confinement inside a low-index 2D dielectric waveguide using metamaterial claddings. Confining a guided wave inside a transparent low index dielectric with arbitrary cross section. The momentum transforming cladding surrounding the core leads to simultaneous total internal reflection and rapid decay of evanescent waves outside the core

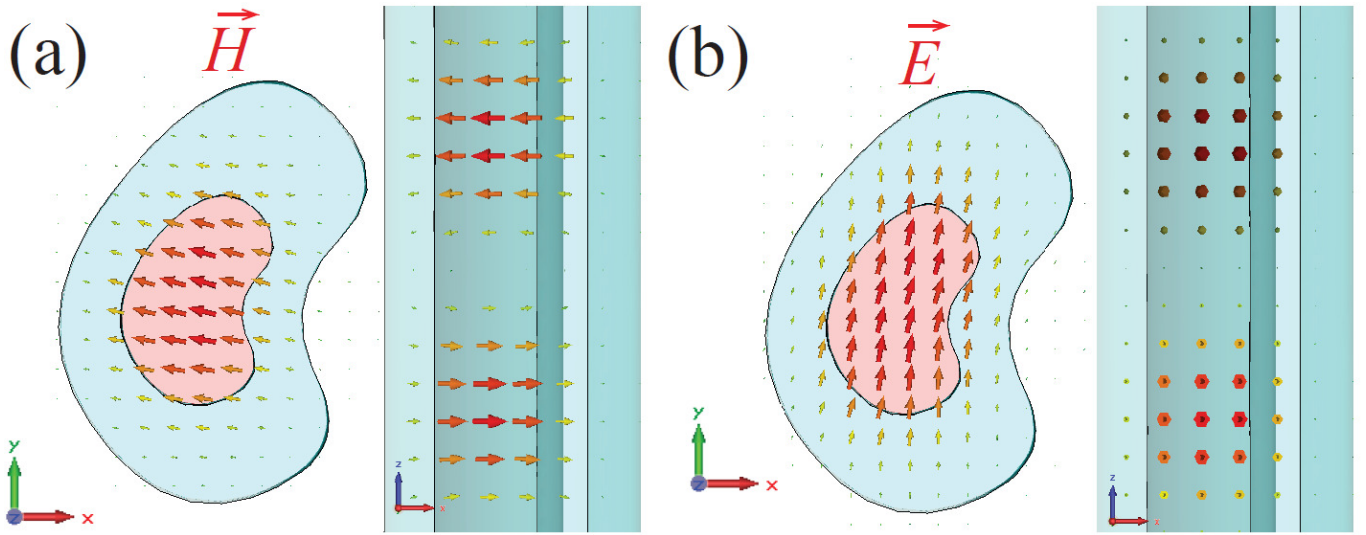

Fig. 3. The simulated (a) magnetic and (b) electric fields across the cross section of the waveguide and along the longitudinal direction. The core consists of a sub-diffraction fiber glass with average radius of $0.1\lambda$ covered by a transformed cladding ($h_x = h_y = 5$ and $h_z = 1.2$) with a size twice as large as the core size. The results show that the electric and magnetic components along the waveguide axis are negligible, so the propagating mode is almost TEM. This is in stark-contrast to the HE11 mode of an optical fiber which has both transverse and longitudinal components of the electric field.

Note that we have decoupled the total internal reflection criterion ($n_1 > n_{2x} = \sqrt{\varepsilon_{2x}}$) from the momentum transformation condition ($n_{2z} \gg 1$) so they can be achieved simultaneously leading to a fundamentally new approach to light confinement in transparent media (Fig. 1). In essence, our non-resonant transparent medium alters the momentum of light entering it and we emphasize that the above set of transformations can be achieved by all-dielectric media ($n>1$).

## 3. Quasi-TEM waveguide

We now apply the momentum transformation to surround an infinitely long glass rod with arbitrarily shaped cross section ($A \ll \lambda^2$). The electromagnetic grid has a finite width and ideally needs to achieve $h_x, h_y \gg 1$ and $h_z < n_{core}$ to allow for the lowest-order mode (HE$_{11}$) to travel inside the glass core and bounce off by total internal reflection but simultaneously decay away rapidly causing sub-diffraction confinement of the mode (Fig 2). This transformation also causes the longitudinal components of fields, in comparison to the transverse ones, to go zero. Indeed, the electric and magnetic fields for the transformed waveguide can be related to the untransformed ones as

$$\frac{E_{x'}}{E_{z'}} = \frac{h_x}{h_z}\frac{E_x}{E_z} = \gamma\frac{E_x}{E_z} \quad and \quad \frac{H_{x'}}{H_{z'}} = \frac{h_x}{h_z}\frac{H_x}{H_z} = \gamma\frac{H_x}{H_z} \tag{5}$$

and due to the large confinement factor ($\gamma$), the longitudinal field components become negligible. Thus the transformed propagating mode is a quasi-TEM mode, and in contrast to conventional waveguides at low-frequencies, it does not need two reflectors or perfect conductor at boundaries. Fig. 3.a and 3.b shows the simulation results of magnetic and electric field vectors, respectively, for a sub-diffraction arbitrarily shaped glass waveguide with average radius of $0.1\lambda$ covered by a transformed cladding ($h_x = h_y = 5$ and $h_z = 1.2$). The simulations have been done by Finite Integration Technique (FIT) commercial software CST Microwave Studio™ [17]. Note that we have used the relaxed condition of $h_z = 1.2$ since the inner medium is glass not air. It can be seen that fields are concentrated in the low-index sub-diffraction dielectric, and they are almost transverse to the propagation direction.

The class of artificial media that lead to these momentum transformations will have $\varepsilon_x, \varepsilon_y < \varepsilon_{glass}$ and $\mu_x, \mu_y < \mu_{glass}$ while $\mu_z, \varepsilon_z \gg 1$. Note that we also have $\varepsilon_x = \mu_x, \varepsilon_y = \mu_y$ and $\varepsilon_z = \mu_z$ thus allowing single mode propagation inspite of the anisotropy. We term this class of artificial media as dual anisotropic giant birefringent metamaterials. Although momentum transformations, unlike conventional TO applications, can be fulfilled by homogenous materials, the cladding must be dual-anisotropic which is very difficult to implement at optical frequencies. However, general dual-anisotropic structures can be implemented at terahertz or microwave frequencies [18].

With a nonmagnetic cladding, we can only transform the electric field momentum in the cladding. However, even this reduced implementation can control the skin depth in the cladding and confine energy inside the core. As we display in the next section, one set of non-magnetic media which can cause the momentum transformation are anisotropic homogenous dielectric materials with $\varepsilon_x = \varepsilon_y < \varepsilon_{core}$ and $\varepsilon_z \gg 1$.

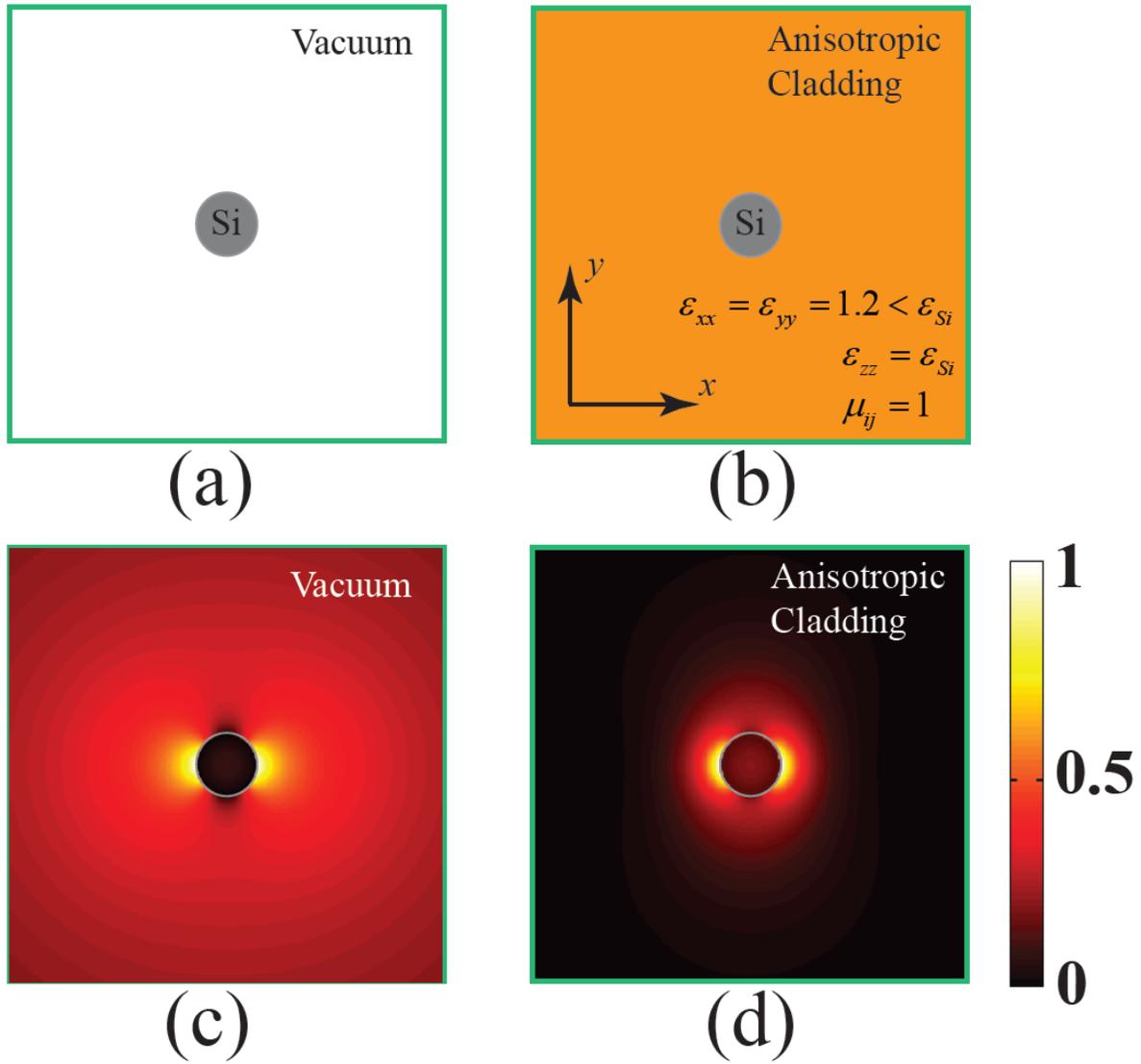

Fig. 4. Is the transparent anisotropic cladding better than vacuum cladding for confinement? (a) It is known that a silicon waveguide in vacuum is the best waveguide for low loss power confinement at optical frequencies. (b) However, we conclusively prove that if the cladding is strongly anisotropic, light confinement can be increased substantially. (c) The x-component of the electric field of $HE_{11}$ mode for silicon waveguide in vacuum. The core radius is $r = 0.07\lambda$. Less than 2% of the power is confined inside the silicon core. (d) The x-component of the electric field of $HE_{11}$ mode for the same waveguide surrounded by an anisotropic cladding ($\varepsilon_x = \varepsilon_x = 1.2 < \varepsilon_{Si}$ and $\varepsilon_z = \varepsilon_{Si} = 12$). The cladding helps to confine up to 30% of the total power inside the core.

## 4. Better than vacuum?

It is commonly believed that to confine light inside dielectric waveguides, we should increase the contrast between indices of the core and the surrounding medium. At optical communication wavelengths, silicon has the highest refractive index among lossless dielectrics. Thus it is widely accepted that a silicon waveguide in vacuum can confine light better than any other lossless waveguide (Fig. 4.a). However, if we cover the silicon core with a transparent anisotropic dielectric as demonstrated in fig. 4.b, the waveguide can confine the first HE mode better than the conventional waveguide. To satisfy relaxed-TIR condition, we must have $\varepsilon_x = \varepsilon_y < \varepsilon_{Si}$ and for strong confinement, $\varepsilon_z$ should be as large as possible (

$\varepsilon_z = \varepsilon_{Si}$ in this case). The x-component of the electric field of the two waveguides is compared in fig. 4.c and d. It is seen that the transparent anisotropic cladding outperforms the vacuum cladding waveguide by a factor of 15 in terms of the mode area. The radius of silicon core for the both waveguides is the same ( $r = 0.07\lambda$ ). The anisotropic cladding causes increased power confinement from less than 2% to 30%.

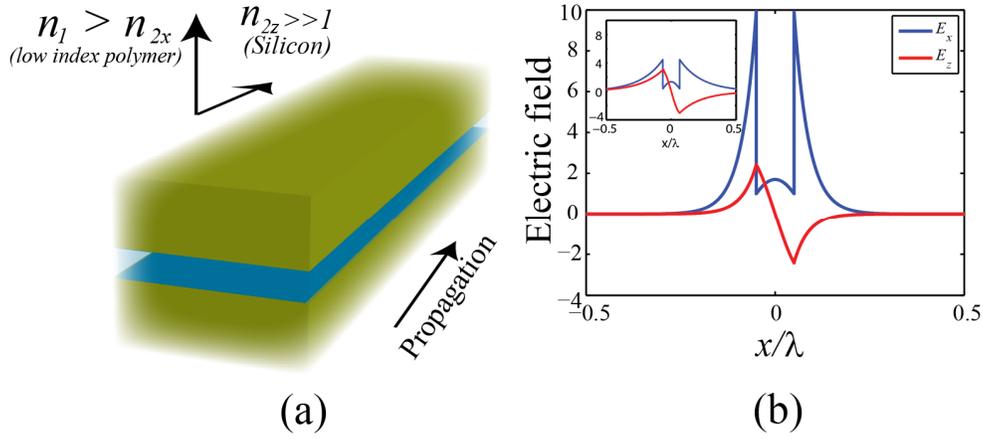

Fig. 5. (a) Schematic representation of a 1D e-skid waveguide. (b) Normalized electric field of a silicon e-skid waveguide with a size of 200 nm operating at 1550 nm. The cladding is anisotropic with $\varepsilon_{2x} = 1.2$ and $\varepsilon_{2z} = 12$. Light decays faster in the anisotropic cladding in comparison with air (inset) which has the lowest refractive index.

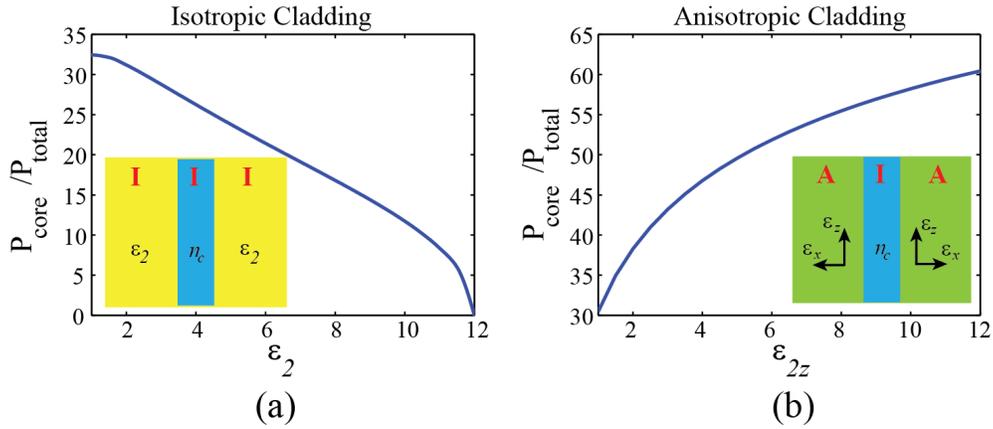

Fig. 6. Power confinement versus cladding index. The core is silicon ( $\varepsilon_c = n_c^2 = 12$ ) with a size of 200 nm operating at $\lambda = 1550 nm$. (a) The cladding is isotropic. If the contrast between the core and cladding index increases, a larger fraction of the total power is confined inside the core. (b) The cladding is anisotropic with $\varepsilon_x = 1.2$. As the anisotropy of the cladding is enhanced, more power is confined inside the core. Thus the conventional waveguide and e-skid waveguide show fundamentally different behavior with increasing cladding index.

## 5. 1D Extreme skin depth waveguides

Skin depth engineering can also be applied to 1D dielectric waveguides leading to sub-diffraction light confinement inside the core. If the slab size is small enough, the fundamental TE and TM modes propagate with no cut-off. These modes in conventional waveguides leak considerably into the cladding and are poorly confined inside the core.

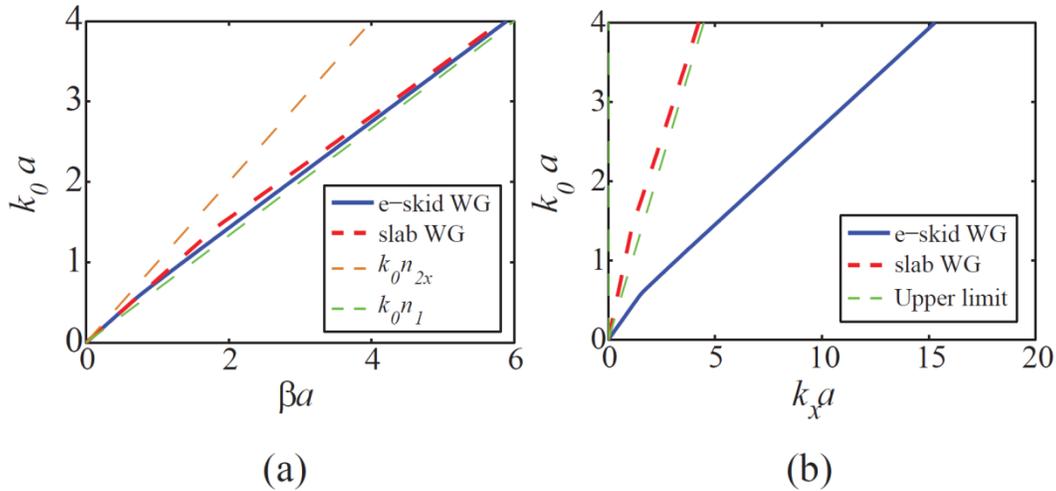

Fig. 7. (a) The propagation constant ($\beta$) dispersion of the first TM mode normalized to the core size ($a$). (b) The decay constant dispersion of the first TM mode in the cladding. The core is glass. The isotropic cladding is air and the anisotropic cladding has a permittivity of $\varepsilon_x = 1.2$ and $\varepsilon_z = 15$. The propagation constant cannot exceed the light line in the core and cladding, but the decay constant in the anisotropic cladding can exceed the upper limit.

We can implement transparent anisotropic metamaterials as the cladding of conventional 1D slab waveguides to decrease the skin depth in the cladding for the first TM mode (fig. 5.a). To allow waveguiding by total internal reflection, the cladding index perpendicular to the propagation direction ($x$) must be less than the core index. We can simultaneously control the skin depth in the cladding by increasing the cladding index parallel to propagation direction ($z$). The electric field profile of an e-skid waveguide and slab waveguide operating at telecommunication wavelength ($\lambda = 1550 nm$) are compared in fig. 5.b. For the slab waveguide, we chose the highest achievable contrast between the core and the cladding index (inset). We see that a considerable amount of power lies outside the waveguide because the core is small compared to the operating wavelength. However, if we use anisotropic cladding, light is strongly confined inside the core. As it is shown in fig. 6, if we decrease the contrast between the core and cladding indices of a conventional slab waveguide, the power confined in the core is significantly lower. However, if we increase the cladding index only in the $z$ direction, we can enhance the confinement. Note that $\varepsilon_{2z}$ can even be larger than $\varepsilon_1$.

We now contrast the propagation characteristics of e-skid waveguides with conventional waveguides emphasizing the key differences.

**A. Propagation constant:**

Fig. 7.a displays the propagation constant dispersion for an e-skid waveguide and a conventional slab waveguide. It is seen that the dispersion is very similar for the two waveguides. Note that the propagation constant cannot be larger than the wave vector in the core ($\beta < k_0 n_{core}$) because the light is guided in the core by total internal reflection for both cases. Thus the sub-diffraction confinement in e-skid waveguides is fundamentally different from surface wave approaches.

**B. Decay constant:**

The key aspect of e-skid waveguides is that the decay constant in the cladding can exceed the maximum value that can be achieved by an isotropic dielectric cladding. The exact value for the decay constant in the cladding ($k_{2x}$) can be calculated from two coupled nonlinear equations [10]. However, if $k_0 a \ll 1$ ($a$ is the core size), $k_{2x}$ can be approximated as:

$$k_{2x} = \frac{1}{\delta_{cladding}} \cong \frac{\varepsilon_{2z}}{\varepsilon_1} a(\varepsilon_1 - \varepsilon_{2x}) k_0^2 \qquad (6)$$

where $\varepsilon_1$ and $[\varepsilon_{2x}\ \varepsilon_{2z}\ \varepsilon_{2z}]$ are the permittivity of the core and cladding, respectively. We see that as the anisotropy in cladding increases ($\varepsilon_{2z} \gg 1$), the mode decays faster in the cladding and the skin depth ($\delta_{cladding}$) decreases. The decay constant dispersion in the cladding is plotted in fig. 7.b. The decay constant in the e-skid cladding is dramatically larger than the decay constant in the isotropic cladding. This means that the skin depth is extremely low and consequently the mode can be strongly confined below the diffraction limit of light inside the core. In the next section, we use three figures of merit to compare the confinement in e-skid waveguides with that in conventional slab waveguides.

## 6. Figures of merit

We use three figures of merit for measuring confinement in 1D e-skid waveguides: mode length, power in the core, and mode width. Here, we show that if the core size is smaller than the skin depth, the confinement in all figures of merit is proportional to the skin depth in the cladding ($\delta_{cladding} = 1/k_{2x}$), where $k_{2x}$ is the decay constant of the first TM mode in the cladding. These three FOMs clearly show that e-skid waveguides exhibit a larger confinement than conventional waveguides.

**A. Mode length:**

Mode length is derived from the concept of mode volume in quantum optics for 1D structures. It is commonly used for plasmonics and slot waveguides. Mode length is defined as the ratio of the total mode energy and mode energy density peak [4] as $L_m = \int_{-\infty}^{\infty} W(x)dx / \max\{W(x)\}$, where $W(x)$ is the electromagnetic energy density. If $k_0 a \ll 1$, it can be approximated as:

$$L_m \cong \frac{2\delta_{cladding} + a(1 + \varepsilon_{2x}/\varepsilon_1)}{1 + \varepsilon_{2x}/\varepsilon_1} \qquad (7)$$

where $k_{2x}$ is determined from (6). The mode length for an e-skid waveguide with glass core and anisotropic cladding of $\varepsilon_x = 1.2$ and $\varepsilon_z = 15$ is plotted in fig. 8.a in comparison with a conventional glass slab waveguide and air cladding. The mode length is normalized to the diffraction limit of light in the core ($\lambda/2n_{core}$). It is clearly seen that the diffraction limit is surpassed due to the extremely small skin depth in the anisotropic cladding. The numerical calculation of the mode length for the e-skid waveguide is also plotted and there is an excellent agreement with the analytical calculations when the core size is small enough. If the core size is smaller than the skin depth, the second term in the numerator vanishes and the mode length becomes proportional to the skin depth.

**B. Mode width:**

Although the mode length is a good measure of confinement for comparing waveguides with similar field profiles, it is not a fair figure of merit to compare different classes of waveguides, since mode length strongly depends on the peak energy density. If the field profile is not uniform, the mode length does not give any information about the size of the mode. For example, in slot-waveguides the energy density peaks in a very tiny gap surrounded by high index dielectrics [8], so the mode length for slot-waveguides achieves sub-diffraction values ( $L_m \sim 0.1\lambda/2n_{core}$ ). However, the mode decays very slowly outside. Thus it cannot be used in dense photonic integrated circuits due to the cross-talk. Berini [19] has defined mode width as a measure of confinement for applications where the size of the mode is important, e.g. photonic integration. Mode width is the width at which the field intensity falls to $1/e$ of the maximum field intensity.

$$\delta_w = a + 2/k_{2x} = a + 2\delta_{cladding} \tag{8}$$

If the skin depth reduces, the mode decays faster in the cladding and the mode width decreases. The mode width of the e-skid waveguide is compared with the conventional slab waveguide in fig. 8.b. The structure is the same as the structure in fig. 7.a and results are normalized to the diffraction limit of light in glass. E-skid waveguides have a smaller mode width and can surpass the diffraction limit.

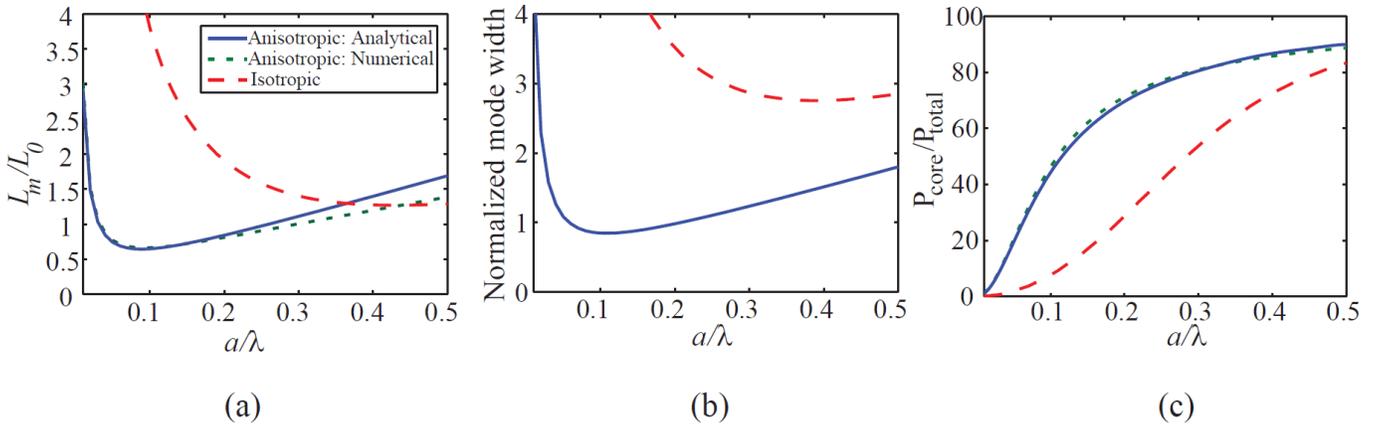

(a)                          (b)                          (c)

Fig. 8. Figures of merit for 1D e-skid waveguides. (a) mode length, (b) mode width, and (c) power confinement. The structure is the same as that in fig. 7.

**C. Power in the core:**

Another figure of merit for confinement is the fraction of power in the core. An ideal confinement is when all of the power is in the core area and the cladding carries no power. If $k_0 a \ll 1$, the ratio of power in the core and power in the cladding can be approximated as:

$$\frac{P_{core}}{P_{cladding}} \cong \frac{a}{2\delta_{cladding}}\left(1+\frac{\varepsilon_{2x}}{\varepsilon_1}\right) \qquad (9)$$

As the skin depth decreases, a larger fraction of the total power is confined inside the core. The ratio of power in the core and total power (in percent) is plotted in fig. 8.c for the e-skid waveguide and slab waveguide. The numerical calculation of the power is also plotted for comparison and an excellent agreement is observed.

## 7. Goos–Hänchen phase shift

The key building block for e-skid photonics are transparent anisotropic metamaterials. Designs for transparent metamaterials utilizing only semiconductors were provided in our previous work [10]. Here, we outline how to verify the concept of extreme skin depth using the Goos–Hänchen phase shift.

Total internal reflection of a beam causes a lateral displacement which is known as Goos–Hänchen phase shift [20]. If the skin depth at total internal reflection decreases, the Goos–Hänchen phase shift decreases as well.

If the entire angular spectrum of the beam is above the critical angle and the center of spectrum is at $\theta_0$, the Goos–Hänchen phase shift can be calculated as [16]:

$$D = -\frac{\lambda}{2\pi}\frac{d\varphi(\theta)}{d\theta}\bigg|_{\theta=\theta_0} \qquad (10)$$

where $\varphi(\theta)$ is the reflection phase of an incident plane wave at the incident angle of $\theta$. The reflection coefficient is $r = H_y^r/H_y^i = (k_{1x}\varepsilon_{2z} - k_{2x}\varepsilon_1)/(k_{1x}\varepsilon_{2z} + k_{2x}\varepsilon_1)$, where $k_{1x}$ and $k_{2x}$ are optical momenta normal to the interface between two dielectrics, $H_y^i$ and $H_y^r$ are total magnetic field of the incident and reflected waves, respectively. The reflection amplitude and phase at the interface of glass and a transparent anisotropic metamaterial ($\varepsilon_{2x} = 1.2$ and $\varepsilon_{2z} = 20$) boundary are compared with that at the interface of glass with an isotropic low-index dielectric in fig. 9.a and b, respectively. Above the critical angle where $k_{2x}$ is imaginary, if $\varepsilon_{2z}$ increases, the reflection phase reduces because the imaginary part of reflection coefficient becomes negligible in comparison with the real part. Thus according to (10), the Goos–Hänchen phase shift decreases as the skin depth reduces. The Goos–Hänchen phase shift of a light beam versus $\varepsilon_{2z}$ is plotted in fig. 9.c. We assume the center of the incident beam's angular spectrum is 1% above the critical angle and all of the angular spectrum components are greater than the critical angle.

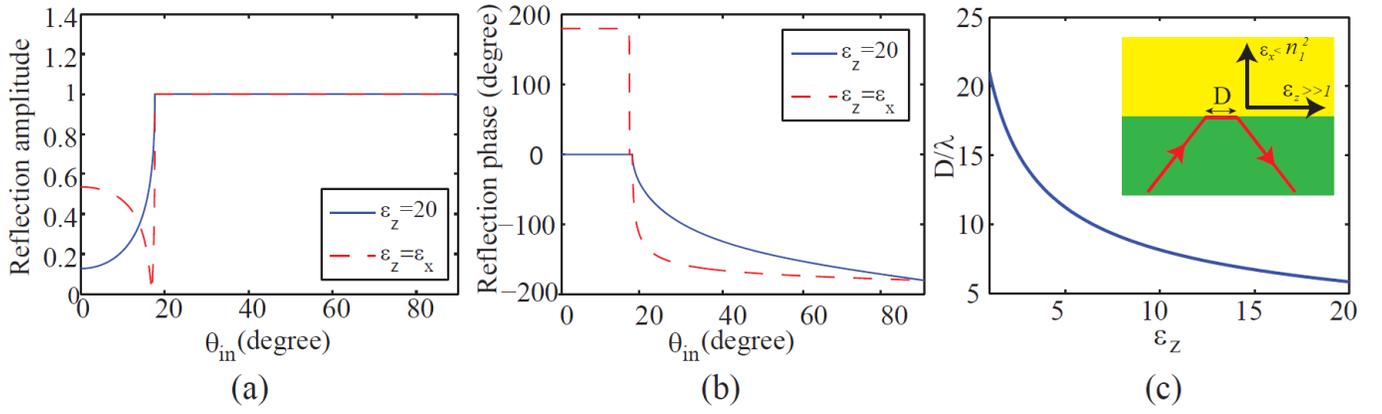

Fig. 9. (a) Reflection amplitude and (b) reflection phase versus the incident angle. The first medium is glass and for the second medium $\varepsilon_x = 1.2$. (c) Goos–Hänchen phase shift versus $\varepsilon_z$ of the second medium. We assume the center of angular spectrum of the incident beam is 1% above the critical angle and all of the angular spectrum components are greater than the critical angle. As the anisotropy increases (i.e. skin depth decreases), the Goos–Hänchen phase shift decrease

## 8. Conclusion

In summary, we have shown that it is possible to reduce the skin depth using transparent anisotropic metamaterials. Three figures of merit were calculated analytically to show e-skid waveguides outperform conventional waveguide in term of light confinement. We showed that the Goos–Hänchen phase shift is reduced in transparent anisotropic metamaterials, a key signature for experimental verification of the skin depth. We also introduced the method of momentum transformations to explain relaxed total internal reflection. Photonic skin-depth engineering can emerge as an important design principle for sub-diffraction optical devices.

## Appendix A: Transforming the momentum of light

Here, we outline the derivation of the dispersion relation when the coordinate system is transformed. The time-harmonic source-free Maxwell's equations in general coordinate system can be written as [13]:

$$\begin{cases} \dfrac{1}{h_i h_j}\left[\dfrac{\partial}{\partial \xi_i}(h_j H_j) - \dfrac{\partial}{\partial \xi_j}(h_i H_i)\right] = -i\omega\varepsilon_k E_k \\ \dfrac{1}{h_i h_j}\left[\dfrac{\partial}{\partial \xi_i}(h_j E_j) - \dfrac{\partial}{\partial \xi_j}(h_i E_i)\right] = +i\omega\mu_k H_k \end{cases} \quad (A1)$$

where $[\xi_i, \xi_j, \xi_k]$ are generalized coordinate components, $[h_i^2, h_j^2, h_k^2]$ are dimensionless Jacobian matrix coefficients, $E_i$ and $H_i$ are electric and magnetic fields vector in $\xi_i$ direction, and $\varepsilon_i$ and $\mu_i$ are tensor components of permittivity and permeability, respectively. If the material parameters and fields are changed according to the Jacobian matrix coefficients, Maxwell's equations are invariant. This can be used to control light propagation also known as transformation optics (TO) theory [13,14]:

$$\varepsilon_i^{new} = \frac{h_j h_k}{h_i} \varepsilon_i^{old}, \quad \mu_i^{new} = \frac{h_j h_k}{h_i} \mu_i^{old}$$
$$E_i^{new} = h_i E_i^{old}, \quad H_i^{new} = h_i J_i^{old}$$
(A2)

Here, we show that when the coordination system is transformed, the dispersion relation which governs optical momentum is also transformed accordingly. This is important for controlling evanescent waves. We first find the plane wave solution (i.e. $E_i = E_{i0} e^{i(k_i \xi_i + k_j \xi_j + k_k \xi_k)}$, $H_i = H_{i0} e^{i(k_i \xi_i + k_j \xi_j + k_k \xi_k)}$) of (A1):

$$\begin{cases} \frac{k_i}{h_i} H_{j0} - \frac{k_j}{h_j} H_{i0} = -i\omega \varepsilon_k E_{k0} \\ \frac{k_i}{h_i} E_{j0} - \frac{k_j}{h_j} E_{i0} = +i\omega \mu_k H_{k0} \end{cases}$$
(A3)

which can be simply written in three sets of independent equations. Assume that the old system is Cartesian composed an isotropic material with refractive index of $n$. The equations in the new coordinate system $(x', y', z')$ must be in the following form that Maxwell's equations are invariant:

$$\begin{bmatrix} n^2 k_0^2 - \frac{k_{y'}^2}{h_y^2} - \frac{k_{z'}^2}{h_z^2} & \frac{k_{x'} k_{y'}}{h_y^2} & \frac{k_{x'} k_{z'}}{h_z^2} \\ \frac{k_{x'} k_{y'}}{h_x^2} & n^2 k_0^2 - \frac{k_{x'}^2}{h_x^2} - \frac{k_{z'}^2}{h_z^2} & \frac{k_{y'} k_{z'}}{h_z^2} \\ \frac{k_{x'} k_{z'}}{h_x^2} & \frac{k_{y'} k_{z'}}{h_y^2} & n^2 k_0^2 - \frac{k_{x'}^2}{h_x^2} - \frac{k_{y'}^2}{h_y^2} \end{bmatrix} \begin{bmatrix} E_{x'0} \\ E_{y'0} \\ E_{z'0} \end{bmatrix} = 0$$
(A4)

The determination of the above matrix must be zero for nonzero fields, which leads to the following dispersion relation:

$$\frac{k_{x'}^2}{h_x^2} + \frac{k_{y'}^2}{h_y^2} + \frac{k_{z'}^2}{h_z^2} = k_0^2$$
(A5)

This equation shows that when the fields are transformed, the isofrequency (dispersion) curve is also transformed and the momentum in the new system is related to the old one as: $k_{x'} = h_x k_x$, and etc. Note that although constitutive parameters are anisotropic in general, the momentum transformation for both polarizations is the same. We term such media as dual anisotropic i.e. $\frac{\varepsilon_i^{new}}{\varepsilon_i^{old}} = \frac{\mu_i^{new}}{\mu_i^{old}}$.

**Acknowledgements**

This work is supported by Natural Sciences and Engineering Research Council of Canada and Helmholtz Alberta Initiative.